\PassOptionsToPackage{colorlinks,bookmarksopen,bookmarksnumbered,citecolor=red,urlcolor=red}{hyperref}
\documentclass[fleqn,10pt]{wlscirep}
\usepackage[utf8]{inputenc}
\usepackage[T1]{fontenc}
\usepackage{enumitem}
\usepackage{moreverb,url}
\usepackage{tabularx}

\usepackage[acronym,shortcuts]{glossaries}
\newacronym{ADLs}{ADLs}{activities of daily living}
\newacronym{iADLs}{iADLs}{instrumental activities of daily living}

\usepackage{xcolor}

\newcommand\revise[1]{\textcolor{black}{{#1}}}

\title{Preferences and strategies for assistive technology use at home by people who are blind}
\title{Assistive technology use in domestic activities by people who are blind}

\author[1,*]{Lily M. Turkstra}
\author[1]{Tanya Bhatia}
\author[1]{Alexa Van Os}
\author[1,2]{Michael Beyeler}
\affil[1]{Department of Psychological \& Brain Sciences, University of California, Santa Barbara}
\affil[2]{Department of Computer Science, University of California, Santa Barbara}
\affil[*]{lturkstra@ucsb.edu}

\begin{abstract}
People who are blind employ unique strategies when performing instrumental activities of daily living (iADLs), often relying on multiple sensory modalities and assistive technologies. While prior research has extensively explored adaptive strategies for outdoor activities like wayfinding and navigation, less emphasis has been placed on the information needs and problem-solving strategies for managing domestic activities. To address this gap, our study presents insights from 16 semi-structured interviews with individuals who are either legally or completely blind, highlighting both the current use and potential future applications of technologies for home-based iADLs.  
\revise{Our findings reveal several underexplored challenges, including the difficulty of locating misplaced objects, a structured problem-solving approach where digital tools are a last resort, and limited awareness of assistive training programs. Participants also faced persistent usability barriers as software updates disrupted accessibility features.}  
Participants utilize a variety of low-tech and high-tech solutions, with tactile labeling systems and digital assistance apps being particularly prevalent. However, \revise{existing assistive technologies often fail to integrate seamlessly with users’ preferred strategies, leading to frustration and underutilization}. Addressing these barriers is crucial for enhancing the adoption of assistive technologies and ultimately improving the quality of life for people who are blind.  
\end{abstract}

\begin{document}

\flushbottom
\maketitle

\thispagestyle{empty}

\section*{Introduction}

People who are blind may face unique challenges when performing instrumental activities of daily living (iADLs), which refer to the daily activities necessary for independent living, such as housekeeping, meal preparation, managing finances and medication, shopping, and transportation \cite{ward_review_1998,jones_analysis_2019,edemekong_activities_2023}.

Key among these are technological accessibility, which influences tasks like finance management and cooking \cite{li_non-visual_2021}; social support networks, essential for activities such as shopping and transportation \cite{kostyra_food_2017}; and economic constraints, which may restrict access to assistive technologies and home modifications \cite{ulldemolins_social_2012}. The design of living spaces \cite{branham_collaborative_2015}, education in orientation and mobility \cite{giudice_blind_2008,brunet_strategies_2018,cassidy_identifying_2017}, age, co-occurring disabilities, and psychological well-being also play critical roles \cite{ward_review_1998,jones_analysis_2019}. 
Each factor contributes to the unique experiences of individuals who are blind, highlighting the need for personalized approaches to support independent living.

\revise{While prior research has extensively examined how blind individuals navigate outdoor spaces using assistive technology \cite{giudice_blind_2008,brunet_strategies_2018,gupta_towards_2020,bandukda_places_2020,hoogsteen_beyond_2022}, fewer studies have investigated their experiences in indoor environments \cite{cassidy_identifying_2017,abdolrahmani_siri_2018,li_indoor_2010,tjan_digital_2005}. Even fewer have focused specifically on domestic iADLs \cite{li_non-visual_2021, jones_analysis_2019,m_storer_deinstitutionalizing_2021}, despite these activities being critical for independent living.}

\revise{Through an anonymous online survey targeted towards individuals with visual impairment who have used a tablet or smartphone, Martiniello \emph{et al.}~\cite{martiniello_exploring_2022} report that a vast majority feel as though mainstream devices, such as smartphones and tablets, are replacing traditional high and low-tech solutions. This perception, held especially by younger users, exemplifies to both users and developers the importance of awareness of the varied use cases to which these tools may apply.}

\revise{Branham and Kane \cite{branham_collaborative_2015} found that the accessibility of domestic environments for blind individuals is closely tied to the competencies and cooperation of sighted co-residents. Storer and Branham \cite{m_storer_deinstitutionalizing_2021} examined the unique strategies blind parents develop for managing childcare and household responsibilities, while Storer et al. \cite{storer_all_2020} explored the decision-making processes behind adopting assistive devices. These studies suggest that home accessibility is influenced not just by technology but also by household dynamics.} 

\revise{Existing research also highlights significant barriers to adopting assistive technologies, including usability challenges, accessibility issues, and difficulties integrating high-tech tools with well-established low-tech strategies \cite{shinohara_observing_2007,abdolrahmani_not_2016,abdolrahmani_siri_2018,abdolrahmani_blind_2020,beyeler_towards_2022,ptito_brain-machine_2021,hu_overview_2019,kasowski_systematic_2023}. Many of these technologies do not work well enough to meet the specific needs of blind users, leading to frustration and abandonment \cite{shinohara_blind_2009,branham_collaborative_2015,cassidy_identifying_2017,jeamwatthanachai_indoor_2019,gupta_towards_2020}.} 
\revise{Despite these valuable contributions, there remains a need for a broader and more nuanced understanding of how blind individuals perform iADLs in the home, how they integrate assistive technologies, and what gaps remain. Our study expands on previous research by systematically investigating these questions through semi-structured interviews with a diverse group of 16 blind individuals.} 

Our research aimed to answer the following questions:
\begin{enumerate}[topsep=2pt, parsep=2pt, itemsep=0pt]
    \item What are the information needs, strategies employed, and challenges experienced in domestic activities by people who are blind?
    \item How frequently and in what ways are tactile and digital tools used in daily household tasks?
    \item What suggestions and considerations can be provided for the development of future assistive technologies for home use?
\end{enumerate}

By answering these questions, we aim to provide a more in-depth understanding of the specific needs that people who are blind encounter in their daily lives and the limitations of existing technology in addressing them. Identifying common strategies that people use to master these tasks may benefit others who are blind, and recognizing challenges faced (even with the use of tactile labeling systems and smart devices) may pinpoint opportunities for the development of future assistive technologies. 

\revise{Unlike previous work that focuses on specific use cases or narrow populations, our study provides a comprehensive view of assistive technology use across a diverse group of blind individuals. Our findings contribute new insights into the interplay between low-tech and high-tech tools, the persistent barriers to technology adoption, and the unmet needs that remain in domestic settings. By identifying these gaps, we provide actionable recommendations for future assistive technology design and training programs.}

\section*{Methods}

\subsection*{Participants}

Our study targeted a diverse cohort of legally blind adults to explore the challenges they encounter in everyday domestic activities. We recruited 16 participants (Table~\ref{tab:participants}), aged 25 to 79, from the greater Santa Barbara, CA area through partnerships with the SB Braille Institute and Blind Fitness, two local organizations for the blind.

Eligibility criteria included being over 18 years old, legally blind, and fluent in English. \revise{There were an even number of participants who identified as female or male}, and the group included both individuals with congenital (four participants) and acquired blindness (twelve participants).
Participants were first diagnosed between 4 and 74 years ago, with causes of blindness including glaucoma, diabetic retinopathy, multiple sclerosis, retinitis pigmentosa, and other conditions.

Participants varied in their living situations, with most living with sighted spouses and one living alone with occasional assistance from a hired helper. This diversity provided a broad perspective on the daily challenges and strategies employed by blind individuals.

All procedures were approved by the Institutional Review Board of the University of California, Santa Barbara \revise{(Protocol No.~7-23-0035)} and adhered to ethical guidelines for human \revise{participants}.  
Informed consent was obtained from all participants, who were explicitly informed of their right to withdraw from the study at any time.  
\revise{Given the visual impairments of our participants, the consent process was adapted to ensure full comprehension and voluntary participation~\cite{wittich_methodological_2023}. Recruitment materials, including the consent form and an informational flyer, were provided in accessible formats ahead of the study. During the consent process, researchers read the form aloud, allowed ample time for questions, and assisted participants in signing using a signature guide and a high-contrast pen.}

\begin{table*}[hb!]
    \centering
    \renewcommand{\arraystretch}{1.2}
    \resizebox{\textwidth}{!}{
    \begin{tabular}{|rrrrrrrr|}
        \hline
         {\bf ID } & {\bf Age} &  {\bf Years blind} & {\bf Degree} & {\bf \revise{Vision profile}} & {\bf Diagnosis} & {\bf Living situation} & \revise{\bf Employment} \\
         \hline
         A & 70--74  & 5--9 & legally blind & \revise{low vision} & glaucoma (congenital) & with blind & \revise{none}\\
         B & 50--54 & 5--9 & completely blind & \revise{NLP} & multiple sclerosis & with sighted & \revise{part-time}\\
         C & 65--69 & 0--4 & legally blind & \revise{NLP/LP} & \revise{glaucoma} \& diabetic retinopathy & with sighted & \revise{none}\\
         D & 75--79  & 50--54 & legally blind & \revise{low vision} &  Doyne honeycomb \revise{choroiditis} & with sighted & \revise{part-time}\\
         E & 70--74 & 15--19 & legally blind & \revise{peripheral loss} & retinitis pigmentosa & with sighted & \revise{none}\\
         F & 65--69 & 65--69 & completely blind & \revise{NLP} &  unknown (congenital) & with sighted & \revise{none}\\
         G & 65--69 & 10--14 & legally blind  & \revise{low vision} & diabetic retinopathy & with sighted & \revise{partial}\\
         H & 70--74 & 70--74 & legally blind & \revise{peripheral loss} & glaucoma (congenital) & with sighted & \revise{part-time}\\
         I & 65--69  & 0--4 & legally blind & \revise{not disclosed} & optic neuropathy & with sighted & \revise{none}\\
         J & 35--39 & 0--4 & legally blind & \revise{NLP/BLP}  & diabetic retinopathy & with sighted & \revise{none}\\
         K & 50--54 & 15--19 & legally blind & \revise{NLP} & diabetic retinopathy & with sighted & \revise{none}\\
         L & 35--39 & 35--39 & legally blind & \revise{LP} &  Leber's amaurosis (congenital) & with sighted& \revise{full}\\
         M & 40--44 & 10--14 & completely blind & \revise{NLP}  & uveitis & with sighted & \revise{none}\\
         N & 30--34 & 10--14 & completely blind & \revise{NLP} & vehicle accident & with sighted & \revise{none}\\
         O & 25--29 & 20--24 & completely blind & \revise{NLP} & retinoblastoma & alone & \revise{full}\\
         P & 65--69 & 45--49 & legally blind & \revise{LP/NLP} & retinitis pigmentosa & with sighted & \revise{part-time}\\
         \hline
    \end{tabular}
    }
    \caption{Demographic characteristics of our interview participants. \revise{Where two values are given, vision profiles refers to light perception in left/right eye; LP: light perception, NLP: no light perception, BLP: bare light perception.}}
    \label{tab:participants}
\end{table*}

\subsection*{Domestic iADLs}

{Instrumental activities of daily living (iADLs) are complex tasks essential for maintaining independence and managing personal affairs.
These activities require more advanced cognitive and motor skills than basic activities of daily living (ADLs), such as eating and dressing.
Although the exact list of iADLs varies across publications, most include home maintenance, meal preparation, managing finances and medication, shopping, and transportation.}

{While extensive research exists on outdoor activities like wayfinding and navigation, this study focuses on domestic iADLs, defined as the subset of iADLs essential for maintaining and managing a household. This list included:}

\begin{itemize}[topsep=2pt, parsep=2pt, itemsep=0pt]
    \item {Housekeeping: Cleaning the house and other tasks that keep the living environment clean and orderly.}
    \item {Laundry: Washing, drying, folding, and putting away clothes.}
    \item {Meal preparation: Planning, preparing, and cooking meals.}
    \item {Shopping for groceries and necessities: Purchasing food, household supplies, and other essential items.}
    \item {Managing finances: Paying bills, budgeting, and handling financial matters related to the household.}
    \item {Managing medication: Organizing and taking prescribed medications as needed.}
    \item {Getting dressed: Selecting appropriate clothing and dressing onseself.}
\end{itemize}

\subsection*{Interview protocol}

Semi-structured interviews were conducted using Microsoft Teams or via an audio call over the phone, moderated by the Student Services Manager at the SB Braille Institute. Each interview lasted between 45 minutes to 1.5 hours, with flexibility for follow-up sessions as needed.
The interview guide covered topics such as daily living routines, specific challenges in performing domestic tasks, utilized strategies and tools, and personal backgrounds.

{We asked participants about each of the above mentioned domestic iADLs to gain a detailed understanding of their strategies for managing these tasks. Specifically, we inquired about the specific examples of their daily routines, the assistive technology they used, and the effectiveness of these technologies. Additionally, we explored what aspects of the technology worked well for them, what challenges or limitations they encountered, and the remediation strategies they employed when faced with difficulties.}

Detailed notes were taken in real time, and interviews were recorded (with participant consent) for subsequent transcription, ensuring accuracy in data capture and analysis.

\subsection*{Data analysis}

Our analytical approach was grounded in an embedded case study design \cite{shinohara_observing_2007}, allowing for a close examination of the qualitative data collected. 
We employed thematic content analysis \cite{rabiee_focus-group_2004} to sift through the interview transcripts, identifying and coding recurring themes and patterns {with regard to barriers to adoption of digital assistive technologies} (similar to Refs.~\cite{jones_analysis_2019, abdolrahmani_siri_2018, aghazadeh_strategies_2021}).

{The data analysis began with verbatim transcription of each interview for accuracy. Two researchers independently performed initial coding, reading through the transcripts and assigning preliminary codes to meaningful segments of text. The researchers then compared and merged their codes to create a comprehensive list of themes, ensuring diverse insights were incorporated.
This final list of themes was used to identify barriers to technology adoption, pain points, and remediation strategies.}

\section*{Results}

{Given the} diverse age, demographic backgrounds, and experiences with blindness among our participants, we began the interviews by asking them to describe a typical day in their lives. 
Most participants adhered to a daily routine centered around {domestic} activities such as housework and cooking, with occasional trips to the post office, bank, or shopping mall.
All participants valued staying active and social, either through daily walks with family or friends or by participating in regular exercise classes. Given our collaboration with the SB Braille Institute and Blind Fitness for recruitment, it was unsurprising that many participants frequently attended art, exercise, and assistive technology tutorial classes.

All participants expressed a strong desire for independence in their daily activities. Fifteen of the sixteen participants did not live alone, with fourteen of them living with a sighted spouse. The one participant who lived alone relied on a hired helper to assist with cleaning and other potentially unsafe tasks around the house.

\subsection*{Technology use for {domestic iADLs}}

\begin{figure*}[t!]
    \centering
    \includegraphics[width=\linewidth]{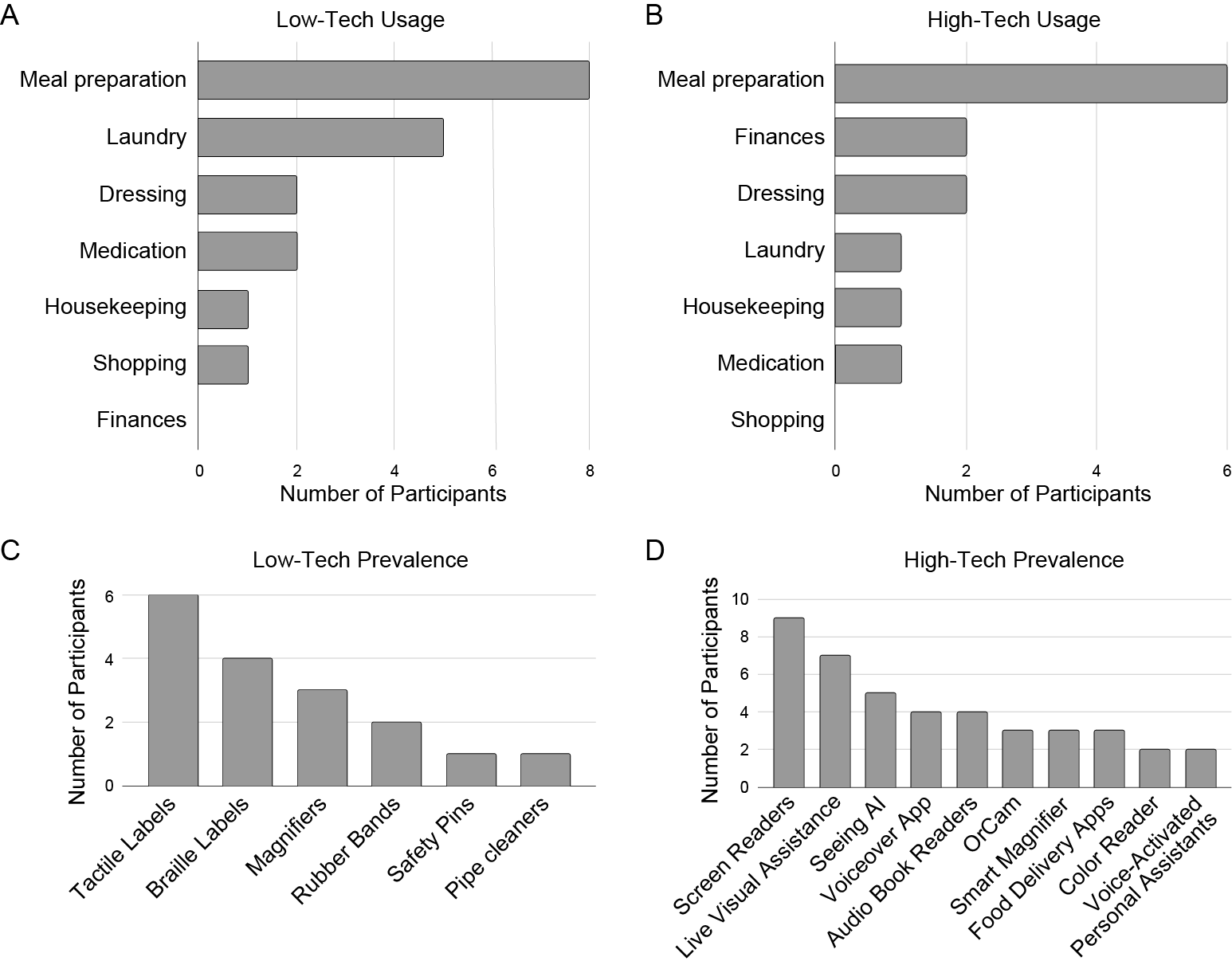}
    \caption{Overview of trends in low- and high-tech usage for {domestic iADLs} from 16 interviewed participants. Panel A and B depict the self-reported frequency of various daily activities for which participants relied on assistive technology. Panels C and D show a breakdown of which low-tech or high-tech solutions were commonly used to support these activities.}
    \label{fig:tech-use}
\end{figure*}

To gain a more systematic understanding of the suitability of various assistive technologies for {domestic iADLs}, we grouped the applications, devices, and services mentioned by our participants into ``low-tech'' solutions (i.e., analog tactile and visual aids) and ``high-tech'' solutions (i.e., digital devices or smartphone applications).
These results are summarized in Fig.~\ref{fig:tech-use}.

In general, participants reported utilizing low-tech solutions for cooking, doing one's laundry, dressing oneself, and cleaning around the home; but not for managing finances (Fig.~\ref{fig:tech-use}A). 
All participants reported relying on one or more tactile labeling systems (Fig.~\ref{fig:tech-use}C); be it using bump dots, textured stickers, rubber bands, pipe cleaners, \revise{braille} Dymo labels, or professional \revise{braille} embossers.
These were often used to label kitchen utensils, cleaning supplies, or important buttons on kitchen appliances and the laundry machine.

In comparison, high-tech apps and services proved useful for all tasks (Fig.~\ref{fig:tech-use}B). They included live visual assistance (e.g., Be My Eyes, Aira) for cooking and cleaning, voice-activated personal assistants for setting timers and automating certain appliances,
voice-over apps (e.g., JAWS) to help with managing finances and reading recipes, and color readers to assist with getting dressed and sorting laundry. 
Two often mentioned products were the Seeing AI app and the OrCam device, which could be used to scan bar codes, identify objects, and recognize currency denominations.

Importantly, low-tech and high-tech solutions were not used in isolation from each other.
For instance, participants may combine tactile labels to label kitchen supplies and appliances, but also rely on digital solutions to scan bar codes and read recipes.
Despite the diversity in preferences and strategies, {our thematic analysis revealed that} our participants engaged in similar heuristics when their iADLs did not go as planned:
 \begin{enumerate}[topsep=2pt, parsep=2pt, itemsep=0pt]
     \item \emph{Prioritize independence:} All of our participants emphasized their desire to be independent. Thus, before relying on external help, they would attempt to perform the task themselves. Many participants expressed frustration with their dependence on spouses and family members, especially for daily activities that ``seem so basic''.
     \item \emph{Tactile aid preference:} Only if their first attempt failed would our participants consider an external aid. When given a choice, they would typically reach for the tactile aid.
     \item \emph{Digital aids as a last resort:} Only if relying on their own senses and any available tactile labels failed would our participants consider a digital aid such as a smartphone app. Participants displayed different preferences for their ``high-tech'' visual aid: whereas many preferred live visual assistance aids (e.g., Be My Eyes, Aira), others preferred AI-based solutions (e.g., Seeing AI, Amazon Echo).
 \end{enumerate}

\revise{Our findings indicate that while assistive technologies are available, their adoption is influenced by user preferences, accessibility barriers, and the individual's degree of prior exposure to rehabilitation services.} 
These results suggest that future digital assistive technologies should focus on addressing issues that tactile solutions cannot solve effectively or completely.

\subsection*{Common pain points \& remediation strategies}

Participants had mixed perceptions, comfort ratings, and motivation levels for technology use at home. However, those who routinely relied on technology reported higher comfort levels and ease of use. Participants consistently praised the availability and accessibility of live visual assistance services, magnifiers, and voice-over tools.

While most everyday activities were reported as doable, they were often tedious and time-consuming. To effectively operate around the home, participants relied on a variety of tactile and digital tools that provided information about the physical layout of their home, the location of objects and furniture, and the state of appliances and household systems. Access to information was further enhanced through tactile or auditory feedback, which helped confirm the successful completion of tasks.

A common strategy among participants, similar to those in Shinohara and Tenenberg~\cite{shinohara_observing_2007}, was the ``brute force backup,'' where participants would try all possible options available as a fallback when other strategies failed. This methodical ``last resort'' heuristic, while potentially exasperating and time-consuming, ensured that tasks were completed and often exposed individuals to new ways of accomplishing various iADLs.

\revise{In addition, we found that many participants were unaware of available training programs and rehabilitation services that could support education and awareness of adopted personal assistive technologies. Instead, they relied heavily on word-of-mouth recommendations and informal peer networks.}

\revise{Our findings reinforce the need for accessible design not just in the development and design of assistive technologies themselves, but also within the infrastructure surrounding their adoption, such as training programs and rehabilitation services. Many of the challenges participants faced were not perpetuated due to a lack of available solutions, but rather to barriers in accessing, learning, and integrating them effectively into their daily routines.}

A brief overview of common pain points and remediation strategies for {the domestic iADLs under study} is given below.

\subsubsection*{Housekeeping}
Cleaning and tidying one's home was a common source of frustration.
The experiences of our participants were mostly consistent with the existing literature \cite{branham_collaborative_2015}.
Participant N highlighted the need to appropriately label and store their cleaning supplies; since these items were not used every day, it was easy to forget where they put them.
Participant I lamented that it was quite difficult to tell whether the floor needed to be swept, stating:
\begin{quote}
    ``It's hard to know if the floor is dirty unless I physically feel something underfoot.''
\end{quote}
Participant M was the only one to report that they actually enjoyed cleaning, sharing:
\begin{quote}
    ``My organizational skills and tidiness have improved since losing my sight. I find it therapeutic.''
\end{quote}

Many participants reported needing to repeatedly sweep or dust their rooms just to make sure they had not missed a spot.
Because of that, most participants relied on the help of a spouse or an external cleaning service.
While this might be an acceptable solution to many, it was perceived to be expensive and to reduce the participant autonomy and independence.

Aside from a few apps that may act as live visual aids, there are little to no existing technologies to support visually impaired and blind individuals in ensuring the cleanliness of their homes.

\subsubsection*{Laundry}
Doing laundry {was reported to be} time-consuming but manageable {for most participants. A significant pain point was the difficulty in using newer washing machines with touchscreens, which made changing or verifying settings challenging.} Participant E reported that: 
\begin{quote}
    ``Doing laundry is fairly easy when no settings need changing. But since we have newer machines, it can be really hard if I do have to fix the settings because the screen is digital.'' 
\end{quote}

To address these challenges, many participants used tactile markers, such as bump dots, to label important buttons. However, identifying stained clothing remains a challenge, as it requires visual confirmation. One participant mentioned:
\begin{quote}
    ``I can do laundry by myself, except for certain stains where I have to ask for help.'' 
\end{quote}

Despite these difficulties, some participants found laundry to be one of their more manageable household tasks, especially when using older machines with dials. Others, like Participant G, employed additional aids such as color coders to assist with sorting laundry.

These findings highlight the need for more accessible and user-friendly designs in modern washing machines.

\subsubsection*{Meal preparation}

Several participants love to cook and use several tools, often embracing the process even if it means ``making a mess'' (Participant O). Most reported being ``slow and meticulous,'' emphasizing the need for an organized kitchen with labeled jars and appliances.

Participant G highlighted the importance of organization:
\begin{quote}
    ``I cook all the meals for my family.
    However, sometimes [my spouse] comes in and accidentally moves something around. Then it takes me a really long time to find it, and no tool can help you with that. It's super frustrating.''
\end{quote}

Common challenges included reading recipes, locating ingredients, mixing and measuring, dealing with hot surfaces, and determining food doneness. Strategies included preparing ingredients before cooking (``mise en place'') and using specialized tools like no-touch thermometers.

Participants who attended cooking classes used heat-resistant gloves, long oven mitts, larger pans, and techniques for gauging temperature safely. Those with residual light perception stressed the importance of appropriate lighting.

Despite utilizing multiple senses and strategies, many found cooking time-consuming and frustrating, leading some to rely on takeout and food delivery services. Participant L summarized: "Cooking tasks are doable but time-consuming and require different tools at each step, making a huge mess and causing frustration."

These findings highlight the need for accessible kitchen tools and technologies to streamline meal preparation and reduce frustration.

\subsubsection*{Shopping for groceries and necessities}

Shopping for groceries and necessities, while technically an outdoor activity, is integral to managing a household. Participants' experiences varied based on their level of vision, familiarity with the store, and availability of assistance.

All participants highlighted the necessity of help for reading labels and finding items, typically by a sighted store clerk or a spouse. Participant F would write a shopping list using \revise{braille} and then read it to a store clerk. Participant I noted the challenge of store layouts changing frequently, complicating navigation even for regular customers.

These findings underscore the need for supportive technologies that enhance navigation and accessibility in both physical and online retail environments.

\subsubsection*{Managing finances}
Managing personal finances can be particularly challenging without vision, as many banks still send paper statements and web-based platforms often lack accessible design features \cite{wentz_exploring_2017}. 
Most participants relied on the help of a spouse or family member to manage tasks like paying bills and filing taxes. For instance, Participant A expressed unease with using screen readers for personal information, saying:
\begin{quote}
    {``I don't trust putting sensitive information online, so I handle finances through phone conversations and rely on my spouse for the rest.''}
\end{quote}

Participants who managed their finances independently, such as Participants B, H, and M, reported various difficulties, from handling checks to accessing online statements due to poor web interfaces. These challenges often required a combination of assistive technologies. Participant H explained:
\begin{quote}
    {``I use a CCTV to enlarge checks, JAWS to navigate eBanking platforms, and Talk Back for the bank's smartphone app, but I still depend on my spouse for critical transactions.''}
\end{quote}

{Other participants maintained balances and collected tax information but collaborated closely with their spouse for final processing.}

{These findings underscore the need for banks and financial institutions to improve the accessibility of their services.}

\subsubsection*{Managing medication}

{Managing medication was another challenging task for participants, necessitating various strategies and assistive technologies. Participant H used an app called Script Talk, which involved the pharmacy attaching a tag to the medication bottle that the app could read, providing an audio version of the label. Participant H explained:}
\begin{quote}
    {``The pharmacy puts a tag on the bottle, and Script Talk reads everything written on it.''}
\end{quote}

{Participant J used bump dots to differentiate medications for day and night use:}
\begin{quote}
    {``I put bump dots on the top of the dresser and on the touch screen of the laundry machine, and I use them to label my medications.''}
\end{quote}

{Participant I highlighted the difficulty of locating dropped medications, saying:}
\begin{quote}
    {``If I drop my medicine, it's like it's gone into a black hole on the white tile floor.''}
\end{quote}
To combat this, they used patterned tablecloths and organized their medications into pill boxes for daily use.

These examples underscore the importance of accessible medication management systems, currently requiring a combination of tactile markers, organizational strategies, and assistive technologies. Enhanced design and support for such tools in pharmacies and healthcare settings could further ease this essential daily task.

\subsubsection*{{Getting dressed}}

Participants generally agreed that getting dressed is not difficult but can be time-consuming, with organization being key. 
Many have developed unique and meticulous systems for sorting their clothes by color, texture, or activity, and use \revise{braille} labels and safety pins to tag important pieces. However, determining if colors are matched or if an outfit looks good can still be challenging.

To mitigate this problem, Participant P got creative:
\begin{quote}
    ``I design outfits the night before or when the clothes come fresh out of the washing machine, then I put them all on the same hanger so I know which ones go together.''
\end{quote}
Others use a color reader, ask their partners for help, or rely on live visual assistance apps like Be My Eyes for the final judgment.

For instance, some participants mentioned using color readers to distinguish clothing colors, while others depended on their partner's input. Participant A said, ``I stay very organized, but I can't tell the difference between colors, so I often ask my spouse for help."

These strategies and tools help mitigate the difficulties associated with getting dressed, highlighting the importance of accessible design in clothing management systems. Enhanced assistive technologies could further improve the independence and confidence of visually impaired individuals in choosing their outfits.

\subsection*{Barriers to adoption of digital assistive technologies {in domestic iADLs}}

Using thematic content analysis (see Methods), we identified several potential barriers to the adoption of new digital technologies for {domestic} iADLs, {briefly summarized below.}

\subsubsection*{Lack of awareness}
Many participants felt overwhelmed by the plethora of accessible tech options, none of which seemed to perfectly fit their needs. 
Participant~A exclaimed:
\begin{quote}
    ``By the time you learn one thing, there is a new thing!''
\end{quote}

The most common ways that participants reported acquiring their technology knowledge were through various organizations, demonstrations, and word of mouth.
However, this knowledge did not always translate to technology use.
The prevalence of difficult-to-use touch screens and voice-over apps often led to frustration, technology abandonment, and a return to reliable tactile tools.
As Participant D noted:
\begin{quote}
 ``I know that there is technology out there to use, but I don't use it because I would rather rely on the tools that I already have and the strategies I have already created that work pretty much every time."     
\end{quote}

{Participants suggested that future tech development should focus on creating comprehensive, user-friendly training programs and support systems. These should include accessible online tutorials, user forums, and regular updates from tech providers to help users stay proficient with new technologies.}

\subsubsection*{Training and support}

Participants lamented the extensive training and support required to become proficient users. Some struggled with basic usage, as Participant G stated:
\begin{quote}
 ``I have a lot of apps on my phone, but never use them because they take too long to learn and too long to even pull up.''
\end{quote}
Many participants also noted that while they appreciated the benefits of technology, they found the steep learning curves daunting. They emphasized the need for training to make technology use more accessible and effective.

{Participants recommended that developers include guided tutorials and contextual help within their apps, specifically tailored to individuals with visual impairments.}

\subsubsection*{Accessibility}

Most participants had experienced difficulties adopting digital technology due to accessibility issues. Updates to software often resulted in changes to the user interface, causing further frustration. Participant feedback highlighted that while technologies like Seeing AI and voice-activated assistants were highly valued, frequent updates and changes could make them hard to keep up with.

{Participants suggested that developers ensure consistent accessibility compliance across all updates and versions of software by regularly testing products with visually impaired users to promptly identify and address any accessibility issues.}

\subsubsection*{Technical issues}
Participants appreciated navigation technologies, AI-based apps, live visual assistance technologies, and voiceover tools once they overcame the learning curve. However, technical issues such as compatibility with other devices and low app accuracy discouraged adoption. Participant K was the only participant without major tech complaints. Moreover, some participants expressed concerns about the experimental nature of some technologies, which added to their reluctance to adopt new tools.

{Participants recommended that developers conduct extensive testing to ensure high accuracy and seamless integration with commonly used technologies and services.}

\subsubsection*{Other barriers}
Other commonly mentioned barriers to adoption included the stigma associated with ``fancy" accessibility aids, their high cost, and concerns about privacy due to increasing data collection. Participants noted that while technology could be helpful, it often required significant financial investment and training to use effectively.

{Participants suggested addressing cost barriers by advocating for subsidized pricing or insurance coverage for assistive technologies. They also recommended enhancing privacy features to protect users' personal data and build trust in these technologies.}

\section*{Discussion}

This study explored the practices, challenges, and strategies of 16 blind individuals in performing domestic iADLs. While most activities were reported as doable, they were often tedious and time-consuming. \revise{However, unlike sighted individuals---who may find these tasks monotonous but generally straightforward---blind individuals must navigate additional complexities, including reliance on tactile and auditory strategies, accessibility barriers in mainstream technology, and a persistent lack of support structures.} Our findings highlight several challenges that are \emph{specific} to blind individuals and are not commonly addressed in the literature. 

\revise{First, participants reported significant difficulty when locating lost, dropped, or misplaced objects—an issue that affects nearly every domestic task, yet lacks widely adopted assistive solutions. Unlike sighted individuals, who can visually scan an area for a missing item, blind individuals must rely on tactile search strategies, memory, or digital assistance tools, none of which are optimized for this challenge. This underscores the need for home-focused accessibility solutions that incorporate real-time spatial awareness and object retrieval.} 

\revise{Second, we found that blind individuals tend to follow a structured problem-solving hierarchy when completing household tasks: (1) attempt the task independently, (2) use a tactile strategy, and (3) resort to digital tools only as a last resort. Unlike prior research that primarily examines technology preferences in isolation, our findings suggest that blind individuals often view digital tools as inefficient or disruptive compared to tactile solutions. This contrasts with sighted users, who typically employ vision-first strategies without hesitation. Designing assistive technologies that integrate seamlessly with existing tactile solutions---rather than attempting to replace them---may improve usability and adoption.} 

\revise{Third, despite the availability of various assistive training programs, our findings indicate a \emph{lack of awareness and accessibility of these resources}. Many participants relied on word-of-mouth rather than formal training, suggesting that existing outreach programs do not always effectively reach the broader blind community. This is in contrast to sighted individuals, who generally have easier access to training for household tasks through mainstream educational and social avenues. Future work should explore ways to bridge this gap, ensuring that assistive training is better integrated into technology adoption efforts.}  

\revise{Fourth, participants frequently expressed frustration with mainstream technologies, particularly when software updates disrupted accessibility features. Unlike sighted users, who can visually adapt to new interface layouts, blind users often face significant usability barriers when accessibility tools are altered or removed. These findings emphasize the need for sustained usability testing with visually impaired users throughout the product lifecycle to ensure long-term accessibility, especially as users perceive a upward trend in mainstream technology adoption and development\cite{martiniello_exploring_2022}.}  

\revise{Finally, while tedious iADLs are not unique to blind individuals, the strategies blind individuals employ---such as reliance on multisensory feedback, structured organization, and systematic problem-solving—represent a distinct cognitive and behavioral approach to household task management. Whereas sighted individuals can visually assess and adjust tasks in real time, blind individuals must pre-plan, memorize layouts, and develop routines that mitigate unpredictability. It is likely that one's demographics, such as current visual profile or age, and daily experiences, whether that be conducting chores around the house or carrying out tasks for one's occupation, impact accessible tool usage and strategies. Though specific strategies and approaches may differ based on the visual abilities of an individual or their living situations, recognizing these adaptations can inform the development of assistive technologies that prioritize efficiency, adaptability, and real-world usability.}  

Addressing these barriers is crucial for increasing the adoption of assistive technologies and improving quality of life for blind individuals. Future digital assistive technologies should \emph{not compete with but complement existing tactile strategies}, focusing on solving problems that low-tech solutions cannot address. This approach aligns with user-centered design principles, emphasizing the importance of sustained engagement with the target population. Continued research, innovation, and informed design will lead to more effective tools tailored to the diverse needs of blind users.

\bibliography{references-r2}

\section*{Acknowledgments}

The authors would like to thank the Santa Barbara Braille Institute and Brianna Pettit, founder of the Blind Fitness organization in Santa Barbara, for their assistance in recruiting participants.
{The authors would also like to thank the study participants for their time.}

{\section*{Data Availability Statement}}

{The datasets used and/or analyzed during the current study are available from the corresponding author upon reasonable request.}

\end{document}